\documentclass[12pt]{article}
\usepackage{graphicx,a4}
\usepackage{amssymb}
\begin{document}

%\draft

% Greek letter abbreviations

\def\al{\alpha}
\def\be{\beta}
\def\ga{\gamma}
\def\de{\delta}
\def\ep{\epsilon}
\def\ze{\zeta}
\def\et{\eta}
\def\th{\theta}
\def\io{\iota}
\def\ka{\kappa}
\def\la{\lambda}
\def\rh{\rho}
\def\si{\sigma}
\def\ta{\tau}
\def\up{\upsilon}
\def\ph{\phi}
\def\ch{\chi}
\def\ps{\psi}
\def\om{\omega}
\def\De{\Delta}
\def\Ga{\Gamma}
\def\Th{\Theta}
\def\La{\Lambda}
\def\Si{\Sigma}
\def\Up{\Upsilon}
\def\Ph{\Phi}
\def\Ch{\Chi}
\def\Ps{\Psi}
\def\Om{\Omega}
\def\varep{\varepsilon}

% Environment abbreviations
\newcommand{\ben}{\begin{equation}}
\newcommand{\een}{\end{equation}}
\newcommand{\bea}{\begin{eqnarray}}
\newcommand{\eea}{\end{eqnarray}}
\newcommand{\ba}{\begin{array}}
\newcommand{\ea}{\end{array}}
\newcommand{\bi}{\begin{itemize}}
\newcommand{\ei}{\end{itemize}}

% Symbols etc.

\def\math{\mathsurround 0pt}
\def\oversim#1#2{\lower.5pt\vbox{\baselineskip0pt \lineskip-.5pt
        \ialign{$\math#1\hfil##\hfil$\crcr#2\crcr{\scriptstyle\sim}\crcr}}}
\def\lap{\mathrel{\mathpalette\oversim {\scriptstyle <}}}
\def\gap{\mathrel{\mathpalette\oversim {\scriptstyle >}}}
\def\dbar{{{\,\mathchar'26}\mkern-9mu\!d\,}}
\def\debar{{{\,\mathchar'26}\mkern-9mu\delta}}
\def\no#1{\mbox{\boldmath $:$} #1 \mbox{\boldmath $:$}}

\def\pa{\partial}
\def\half{\frac{1}{2}}
\def\na{\nabla}
\def\ap{\approx}
\def\vp{\varphi}
\def\pt{\propto}
\def\tr{\mathop{\rm tr}\nolimits}
\newcommand{\ket}[1]{|#1\rangle}
\newcommand{\bra}[1]{\langle#1|}
\newcommand{\braket}[2]{\langle#1|#2\rangle}
\newcommand{\sla}[1]{\not\!#1}
\def\omk{{\omega_{\bk}}}
\def\omkp{{\omega_{\bk'}}}
\def\omp{\omega_{\mathbf{p}}}
\def\id{{\mathsf 1}}
\def\prll{\parallel}
\newcommand{\norm}[1]{\frac{1}{\sqrt{2\om_{#1}}}}

\newcommand{\preprintno}[1]
{\vspace{-2cm}{\normalsize\begin{flushright}#1\end{flushright}}\vspace{1cm}}

\title{\preprintno{{\bf NORDITA-02-42 AP}\\{\bf ULB-TH/02-22}} Photon-Axion mixing in an inhomogeneous universe}
\author{Mattias Christensson\thanks{E-mail: mattias@nordita.dk}\\
{\em Nordita}\\
{\em Blegdamsvej 17, DK-2100 Copenhagen, Denmark }\\
\\
Malcolm Fairbairn\thanks{E-mail: mfairbai@ulb.ac.be}\\
{\em Service de Physique Th\'eorique, CP225}\\
{\em Universit\'e Libre de Bruxelles, B-1050 Brussels, Belgium}}

\date{April 2003}

\maketitle
\begin{abstract}
\noindent
We consider the dimming of photons from high redshift type 1a supernovae through mixing with a pseudoscalar axion field in the intergalactic medium.  We model the electron density using a log-normal probability distribution and assume frozen in magnetic fields.  Assuming the magnetic fields are produced in the early universe we are unable to obtain sufficient dimming in order to explain the apparent acceleration without violating the bounds on the frequency dependence of the dimming.  We also show that any axion mixing leading to a reduction in optical luminosities would also lead to a significant reduction in the polarisation of UV light from intermediate redshift objects which may be detected in the future.
\end{abstract}

\vskip.5pc

%%%%%%%%%%%%%%%%%%%%%%%%%%%%%%%%%%%%%%%%%%%%%%%%%%%%%%

\section{Introduction}

At the time of writing, the present understanding of the energy content of the universe appears to be converging towards a standard model.  Observations of the cosmic microwave background radiation seem to suggest that the total amount of energy in the universe is such that there is zero spatial curvature on the largest observable scales \cite{maxboom}.  At the same time, observations of the distribution of matter at smaller scales give us information about galaxy clustering and the time of matter-radiation equality, and tell us that the fraction of this total energy which is in the form of matter is only about $30\%$ \cite{2df}. This is much larger than the fraction of the total energy which can be in the form of baryonic matter \cite{pagel}, and together with observations of the rotation curves of galaxies and velocity dispersions of galaxies within clusters, has lead to the concept of dark matter.  However, there still remains the question of the remaining unaccounted for $\sim 70\%$ of the total energy density.

Although the data set is effectively still quite small, observations of type 1a supernovae seem to have cast some light on this missing energy \cite{sn,sndim}.  The apparent dimness of these supernovae at high redshift suggests that there is a non-zero cosmological constant which makes up the missing energy or at least something which acts like one.  If we are to interpret this cosmological constant as some energy field then we find that its energy density $\rho\sim (10^{-3} eV)^{4}$ is far away from any of the energy scales (i.e. the Planck scale or the scale of supersymmetry breaking) one would naively expect such a field to have if it were non-zero.  It is also not clear how such a small energy scale could be stable to quantum corrections.

Recently it has been suggested that the dimness of high redshift supernovae is not due to the accelerated expansion of the universe, but rather due to mixing between the photons coming from these objects and a pseudo-scalar axion field \cite{Csaki:2001yk}.  The Lagrangian density of the photon-axion system is given by
\ben
\mathcal{L}=-\frac{1}{2}(\partial^\mu a\partial_\mu a+m^2_{a}a^2) +\frac{a}{M_a}F_{\mu\nu}\widetilde F^{\mu\nu}-\frac{1}{4}F_{\mu\nu}F^{\mu\nu}
\een
where $F_{\mu\nu}$ is the electromagnetic kinetic term and $\widetilde F_{\mu\nu}$ its dual, $a$ the axion field, $m_a$ the axion mass and $M_a$ is its coupling to the photon field.  The coupling of the photon and axion fields in this way means that a photon has a finite probability of mixing with its opposite polarisation and with the axion in the presence of an external magnetic field \cite{raffelt}, such as the intergalactic magnetic field.   If the magnetic field is distributed randomly along the path of the light this will lead to a gradual reduction in luminosity until only 2/3 of the initial energy of the flux remains in the form of photons.  

If the axion mass and coupling scales are chosen to be $m_a\sim 10^{-16}$eV and $M_a\sim 4\times 10^{11}$GeV respectively \cite{Csaki:2001yk}, the fractional amount of dimming will not vary much over the optical range where the supernovae are observed.  This might enable one to consider models where the extra energy component of the universe is not a cosmological constant but a fluid with a different equation of state, e.g. a string network $\cite{bucher}$.  The first paper suggesting such a mixing broke the line of sight up into cells within each of which the magnetic field was oriented in a random direction \cite{Csaki:2001yk}.  Later papers included the effect of a redshift varying magnetic field and the presence of intergalactic electrons \cite{Deffayet:2001pc}\cite{Mortsell:2002dd} (see also \cite{Erlich:2001iq}).  If the electron density is assumed to be approximately uniformly distributed in the intergalactic medium with the value $\tilde{n}_{e}=1.8\cdot 10^{-7} \rm cm^{-3}$ the oscillations will strongly depend on the energy of the photons leading to possible conflict with supernovae spectral observations \cite{sndim}. 
However, as pointed out in \cite{Csaki:2001jk}, the intergalactic medium is not uniform and many regions will have a electron density much below the mean density.  This, according to \cite{Csaki:2001jk}, will render the energy dependence unobservable at present. 

In this paper we study the effects of the variation of the free electron density in the plasma on the photon-axion mixing. Due to the large electrical conductivity of the intergalactic medium the magnetic field can be considered frozen into the plasma. The magnitude of the magnetic field will therefore depend on the electron density as $B\propto n_{e}^{2/3}$. In regions with a low electron density the mixing probability will be suppressed because in these regions the magnetic field is reduced. Therefore, when both electron density and magnetic fields are allowed to fluctuate, the total mixing probability for a path along the line of sight is sensitive to regions where fluctuations have produced an electron density high above the mean density. It is therefore important to take fluctuations in both the electron density and the magnetic field into account when evaluating the total mixing probability. 

Throughout this paper we have assumed that the intergalactic magnetic fields were created in the early universe, i.e. at much higher redshifts than any of the objects we will be considering.  There are some models where the intergalactic magnetic fields are produced at much later epochs.  For example it is possible that super massive black holes may be responsible for the amplification of magnetic fields in the intergalactic medium at late times \cite{colgate}.  Different results would be obtained if one were to consider such models. 

\section{Simulations}

In this section we will describe the equations and assumptions we have used in simulating the passage of photons through the intergalactic medium.  We define the background cosmology, describe the photon-axion mixing and then discuss the probability distribution used to model the electron density and magnetic fields.

\subsection{String network cosmology}

We use the standard notation for each component of stress energy in the universe such that the density of component $x$ is given by $\Omega_x=\rho_x/\rho_{crit}$ where $\rho_{crit}=3H_0^2/8\pi G$.  The equation of state of the component $x$ is denoted $\omega_x$ $(P_x=\omega_x \rho_x)$.  The model in $\cite{Csaki:2001yk}$ assumes that the overall density of matter $\Omega_M\sim 0.3$ as suggested by observation, but that the remaining dark energy takes the form of a string network with a different equation of state to that of a cosmological constant \cite{bucher}.  The energy budget is therefore divided up as
\begin{eqnarray}
\Omega_M=0.3\qquad && \qquad \omega_M=0\nonumber\\
\Omega_S=0.7\qquad && \qquad \omega_S=-1/3
\label{string}
\end{eqnarray}
where the energy density of the string network is denoted $\Omega_S$.  The energy density of the string network allows the universe to possess a flat geometry whilst the equation of state of the string component means the Friedman acceleration equation is identical to that of an open matter dominated universe.

The distance $l$ travelled by a photon reaching us today emitted at a redshift $z_e$ within this cosmology is
\begin{eqnarray}
l(z_e)&=&\frac{c}{H_0}\int_{0}^{z_e}\frac{dz}{(1+z)^2(\sum_i\Omega_i(1+z)^{1+3\omega_i})^{1/2}}\nonumber\\
&=&\frac{c}{H_0}\int_{0}^{z_e}\frac{dz}{(1+z)^2(1+0.3z)^{1/2}}.
\end{eqnarray}
Supernova results suggest an $\Omega_M=0.3$, $\Omega_\Lambda=0.7$ universe and one can evaluate the corresponding equations for such a cosmology.  In order for the luminosity distance of such a universe to be mimicked by dimming in a string network universe as described in equation (\ref{string}), one would require a dimming at redshifts of $z\sim 0.5$ and $1$ of about $12\%$ or $15\%$ respectively.  

In order to see if it is possible to obtain such a dimming we need to investigate the photon-axion mixing.

\subsection{Photon-axion mixing}

We need to investigate how an ensemble of photons of different polarisations will evolve in the presence of external electro-magnetic and axion fields.  In order to see fully how the flux and polarisation coming from a source will change we will need to start with an initial density matrix which encodes the information about the polarisation of the source.  However, we will introduce the mixing equations using normal vector potentials.

Our analysis of the mixing follows closely the analysis of \cite{Deffayet:2001pc} and \cite{Mortsell:2002dd}.  Like them we ignore the effects of vacuum polarisation \cite{Adler:1971wn} and the birefringence of the intergalactic medium in magnetic fields.  

We write $A_{\perp}$ and $A_{\parallel}$ for the polarisations of the photon perpendicular and parallel to the magnetic field respectively.  Then since the refractive index is close to unity we can linearise the equations of motion for the axion $a$ and the two polarisations of the photon
\begin{equation}
\left(\omega-i\partial_t+{\cal M}\right)\left[\begin{array}{c}A_{\perp}\\A_{\parallel}\\a\end{array}\right]=0
\end{equation}
where $\omega$ is the energy of the photon and the mixing matrix ${\cal M}$ is given by
\ben  \label{mixmat}
{\cal M}\equiv\left(
\begin{array}{ccccccccc}
\Delta_{p}&0&0\\
0&\Delta_{p}&\Delta_{M}\\
0&\Delta_{M}&\Delta_{m}
\end{array}
\right).
\een
Here the quantities $\Delta_M$, $\Delta_{p}$ and $\Delta_m$ are defined as
\begin{eqnarray}  
\frac{\Delta_M}{1\,{\rm cm}^{-1}}&=&  
       2\times10^{-26}\left(\frac{B_\perp}{10^{-9}\,\rm G}\right)  
       \left(\frac{10^{11}\,\rm GeV}{M_a}\right)  \nonumber
\\  
\frac{\Delta_m}{1\,{\rm cm}^{-1}}&=&  
       -2.5 \times 10^{-28} \left(\frac{m_a}{10^{-16} 
        {\rm eV}}\right)^2 \left(\frac{1 {\rm eV}}{\omega} 
        \right) \nonumber
\\  
\frac{\Delta_{\rm p}}{1\,{\rm cm}^{-1}}&=&  
        -3.6\times10^{-24}\left(\frac{1\,\rm eV}{\omega}  
        \right)\left(\frac{n_e}{10^{-7} \nonumber
        \,{\rm cm}^{-3}}\right).
\label{deltanum}  
\end{eqnarray}  
where $B_\perp$ is the magnetic field strength perpendicular to the direction of propagation of the photon. 

The eigenvalues of the mixing matrix (\ref{mixmat}) are given by 
\ben  \label{eig_1}
\lambda_{1}=\Delta_{p} \qquad{\rm and}\qquad
\lambda_{2,3}=\frac{(\Delta_{p}+\Delta_{m})}{2} \pm
\left(\Delta_{M}^{2}+\frac{\Delta_{p}^{2}}{4}+\frac{\Delta_{m}^{2}}{4}-\frac{\Delta_{p}\Delta_{m}}{2}\right)^{1/2}
\een
where $\lambda_1$ is trivial since the photon can only mix with the polarisation state of the photon parallel to the magnetic field.  The mixing angle which diagonalises the bottom right hand four components of (\ref{mixmat}) is 
\ben  \label{theta}
 \tan2\theta\equiv2\frac{\Delta_M}{\Delta_{p}-\Delta_m}.
\een
We name the rotation associated with this mixing angle $U_{1}$
\ben  \label{rot_to_prim}
U_{1}=
\left(
\begin{array}{ccccccccc}
1&0&0\\
0&\cos\theta&\sin\theta\\
0&-\sin\theta&\cos\theta
\end{array}
\right)
\een
so that we can write the evolution equation for the three states in the form 
\ben  \label{evolution_parallel}
\left\vert\begin{array}{c}A_{\perp}(t)\\A_{\parallel}(t)\\a(t)\end{array}\right\rangle=
U_{1}^{\dagger} D U_{1}
\left\vert\begin{array}{c}A_{\perp}(0)\\A_{\parallel}(0)\\a(0)\end{array}\right\rangle.
\een
Here the matrix $D$ is the diagonal matrix consisting of the propagating eigenstates  
\ben  \label{diagonal}
D = \rm{diag}(e^{-i(\omega+\lambda_1)t},e^{-i(\omega+\lambda_2)t},e^{-i(\omega+\lambda_3)t}).
\een
and equation (\ref{evolution_parallel}) gives the evolution of the $A_\parallel$ and $a$ amplitudes over a single domain.  

In our analysis, the photon/axion states pass through many regions with randomly orientated magnetic fields.  Because of this, $A_{\parallel}$ in one region or cell is not the same as $A_{\parallel}$ in the next and we have to perform an additional rotation within each cell to take this into account.  
It is convenient to express the evolution directly in terms of the $A_{x}$, $A_{y}$ where $x$ and $y$ are fixed directions perpendicular to the direction of propagation. 

We use the name $U_{2}$ for the rotation between the states $A_{x,y}$ and $A_{\parallel,\perp}$ 
\ben  \label{rot_to_x_and_y}
\left[\begin{array}{c}A_{\perp}\\A_{\parallel}\\a\end{array}\right]=
U_{2}\left[\begin{array}{c}A_{x}\\A_{y}\\a\end{array}\right]=
\left(
\begin{array}{ccccccccc}
\cos\psi&-\sin\psi&0\\
\sin\psi&\cos\psi&0\\
0&0&1
\end{array}
\right)
\left[\begin{array}{c}A_{x}\\A_{y}\\a\end{array}\right]
\een
where $\psi$ is the angle between the $A_{\parallel}$ and $A_{y}$ components.
We can then write 
\ben  \label{evolution_x_y}
\left\vert\begin{array}{c}A_{x}(t)\\A_{y}(t)\\a(t)\end{array}\right\rangle=
U_{2}^{\dagger} U_{1}^{\dagger} D U_{1} U_{2}
\left\vert\begin{array}{c}A_{x}(0)\\A_{y}(0)\\a(0)\end{array}\right\rangle
\een
and we introduce the notation for this matrix 
\ben  \label{moster_matrix}
T = U_{2}^{\dagger} U_{1}^{\dagger} D U_{1} U_{2}.
\een
Now the unitary matrix $T$ is a function of $n_e$, $B_{\perp}$ and $\psi$ so it is different for each cell.  We therefore call the $T$ corresponding to the $n$th cell $T_n$, the time of flight after $n$ cells $t_{n}$ and the photon or axion states after $n$ cells $A(t_n)$ and $a(t_n)$ respectively.  We can then write the evolution after $n$ domains as
\ben
\left\vert\begin{array}{c}A_{x}(t_{n})\\A_{y}(t_{n})\\a(t_{n})\end{array}\right\rangle=
T_{n}
\left\vert\begin{array}{c}A_{x}(t_{n-1})\\A_{y}(t_{n-1})\\a(t_{n-1})\end{array}\right\rangle=
T_{n}T_{n-1} ... T_{1}
\left\vert\begin{array}{c}A_{x}(t_{0})\\A_{y}(t_{0})\\a(t_{0})\end{array}\right\rangle.
\een
This shows that in order to work out the effect of mixing in each of the domains along the path of the photon/axion states we can simply multiply the matrices $T_n$ from each domain along that path.  We obtain the total amount of axions and photons after a time $t_n$ by calculating the total conversion probability at the end of the path.

Strictly speaking, this approach is only valid for a pure initial state.  For unpolarised states where the initial state vector has contributions from two perpendicular polarisations one should evolve the full density matrix keeping track of the off-diagonal elements.  However, since the magnetic field direction and strength is randomly generated in each cell, the information in the off-diagonal parts of the density matrix will not be important in the determination of the expected amount of flux and polarisation after a certain time. 

We have verified this in our simulations by tracking the evolution of the full density matrix $\rho$ using
\ben
i\partial_t\rho=\frac{1}{2\omega}\left[M,\rho\right].
\een
We can use the same matrices introduced earlier
\ben
\rho(t_{n}) = T_{n}\;\rho(t_{n-1})\;T^{\dagger}_{n}
\een
so that the density matrix $\rho_{final}$ which gives rise to the probabilities where the wavefunction is broken down, i.e. where the photons are actually observed by telescopes on earth, is given in terms of the emitted state at the source $\rho_{initial}$ by
\ben
\rho_{final}=T_{total}\; \rho_{initial}\;T^{\dagger}_{total}
\een
where $T_{total}=T_{n}T_{n-1}...T_{1}$.

For a given magnetic field strength, the mixing probability $P_{\rm cell}$ increases with decreasing density.  Also, the energy dependence of the dimming becomes smaller with decreasing density.  So the total amount of dimming along each line of sight depends critically upon the behaviour of both the magnetic field and the electron density along each line of sight.  This is what we turn to next.

\subsection{Distribution of electrons and magnetic field}
Much of our treatment of the distribution of electrons will follow the investigation into cosmological magnetic fields of Blasi et al \cite{blasi}.  The high electrical conductivity of the intergalactic medium allows one to assume that the number density of electrons traces out the cosmic matter distribution. Therefore we need to estimate the spectrum of density perturbations at the relatively low redshifts at which type 1a supernovae have been observed.

An initially Gaussian spectrum of density perturbation will evolve into a log-normal probability distribution of the form \cite{lognormal}
\begin{equation} 
P(\delta) d\delta = {1\over {\sqrt{2\pi} \sigma (1+\delta)}} {\rm \, exp} \left({-[\,{\rm ln}(1+\delta) + \sigma^2/2]^2\over {2\sigma^2}}\right) \, d\delta \ . \label{Pdelta} 
\end{equation} 
where the density contrast $\delta=\delta\rho/\rho$.  The parameter $\sigma$ has been calculated numerically using simulations of cosmologies with and without a cosmological constant \cite{bi}.  We are not aware of such a calculation for a string network cosmology but since the variation between the sigma for a flat universe with or an open universe without a cosmological constant is of order 20\% \cite{bi}, we proceed using the expression for the $\Omega_m=0.3,\Omega_\Lambda=0$ universe:

\begin{equation} 
\sigma(z) = 0.08096 + 5.3869 (1+z)^{-1} - 4.21123 (1+z)^{-2} + 1.4433 (1+z)^{-3}.
\end{equation}
One should keep in mind that with this density distribution, most cells are under-dense relative to the mean by a factor of about $\sim 10$.

Since the number density of the electrons traces that of matter, and the average number density of electrons goes as the third power of the size of the universe, we obtain the relationship
\ben 
n_e(z)=(1+\delta(z))\,(1+z)^3\,\tilde{n}_e(0).
\label{n}
\een
This quantity is calculated in each cell, taken to have the size of the Jeans length, along the line of sight using the the probability distribution Eq. (\ref{Pdelta}). 
For the magnetic field, being frozen into the medium, we have the relation
\ben
B_{\perp}(z)=(1+\delta(z))^{\frac{2}{3}}\,(1+z)^2\,\tilde{B}(0)\, \rm sin\psi
\label{b}
\een
where $\psi$ is the angle between the magnetic field and the line of sight.  Since there is more solid angle perpendicular to the line of sight than along it, we took into account the appropriate trigonometric weighting when randomly generating $\psi$ for each cell.  Many uncertainties persist regarding the structure of intergalactic magnetic fields \cite{Kro94}, and their origin is a much debated topic \cite{Kro94}\cite{GraRub}. Here we simply assume that the magnetic fields were created before any of the redshifts that we are interested in. Also, we approximate the coherence scale of the magnetic field to be of the order of the Jeans length. 

Now we have set up the background cosmology, defined the equations for mixing as a function of electron density and magnetic field and determined how these two components behave with redshift, we proceed with the simulations.

\section{Results}
We performed simulations to calculate the effect of axion dimming upon light from high redshift objects based upon the cosmological parameters discussed in the previous section.  We also investigated the energy dependence of this dimming and the effect of the mixing upon the polarisation of distant galaxies.

\subsection{Photon Dimming}

If we consider some flux of light travelling through a cell with a certain probability of mixing, we need to see what proportion of that flux will convert into axions at the end of its path between source and detector.   As mentioned earlier, since the coherence of the magnetic field is lost from cell to cell have taken the orientation of the magnetic field in each cell to be random, implying that axions will on average mix into photons with the same probability.  If we use $f$ to denote the fraction of the original flux of light from some source which remains in the form of photons we can find the value of $f$ after some time $t_{n}$ in terms of the mixing matrix $\rho(t_n)$
\begin{equation}
f(t_n)=\frac{\Sigma_{i=1,2}\rho_{ii}(t_n)}{\Sigma_{j=1,2,3}\rho_{jj}(t_n)}
\label{f}
\end{equation}
which comes about because there are two polarisations of the photon but only one axion.  We simulated random lines of sight out to redshifts of $z=0.5$ and $z=1$ since these are the redshifts around which the type 1a supernova observations are concentrated \cite{sn,sndim}.  

The mass-density within each cell was randomly generated with weighting from the density probability distribution (\ref{Pdelta}), then equations ($\ref{b}$) and ($\ref{n}$) were used to obtain the strength of the magnetic field and the electron density.  The size of each cell was set to the jeans length corresponding to the density of that cell.  We set the overall magnetic field strength and electron density by defining its average value at $z=0$.  The electron density at $z=0$ is fixed to be $\tilde{n}_e(z=0)=1.8\times 10^{-7} \rm cm^{-3}$ by observations of the primordial abundance of light elements whereas we are have more freedom in choosing our value for the magnetic field.  Since the effect of changing the photon axion coupling $M_a$ and the magnetic field leads to a degeneracy in our results, we instead choose to use define a dimming parameter $\eta$ as
\begin{equation}
\eta=2\times 10^{-26}\left(\frac{\tilde{B}(z=0)}{10^{-9}G}\right)\left(\frac{10^{11}{\rm{GeV}}}{M_a}\right) \rm cm^{-1}.
\end{equation}
\begin{figure}
\begin{tabular}{ccc}
\includegraphics[height=6cm,width=7.5cm]{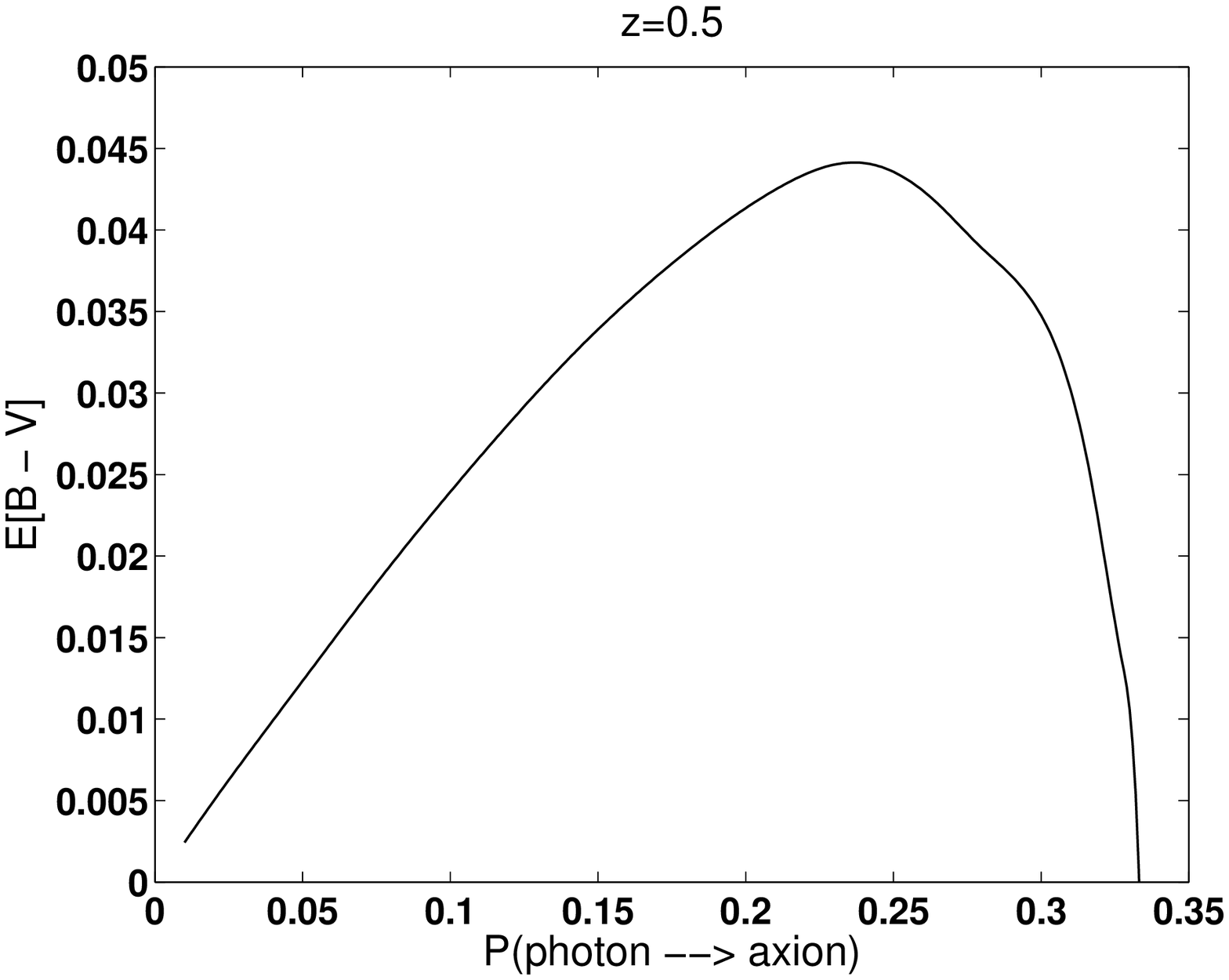} 
&\hspace{1.0cm}&
\includegraphics[height=6cm,width=7.5cm]{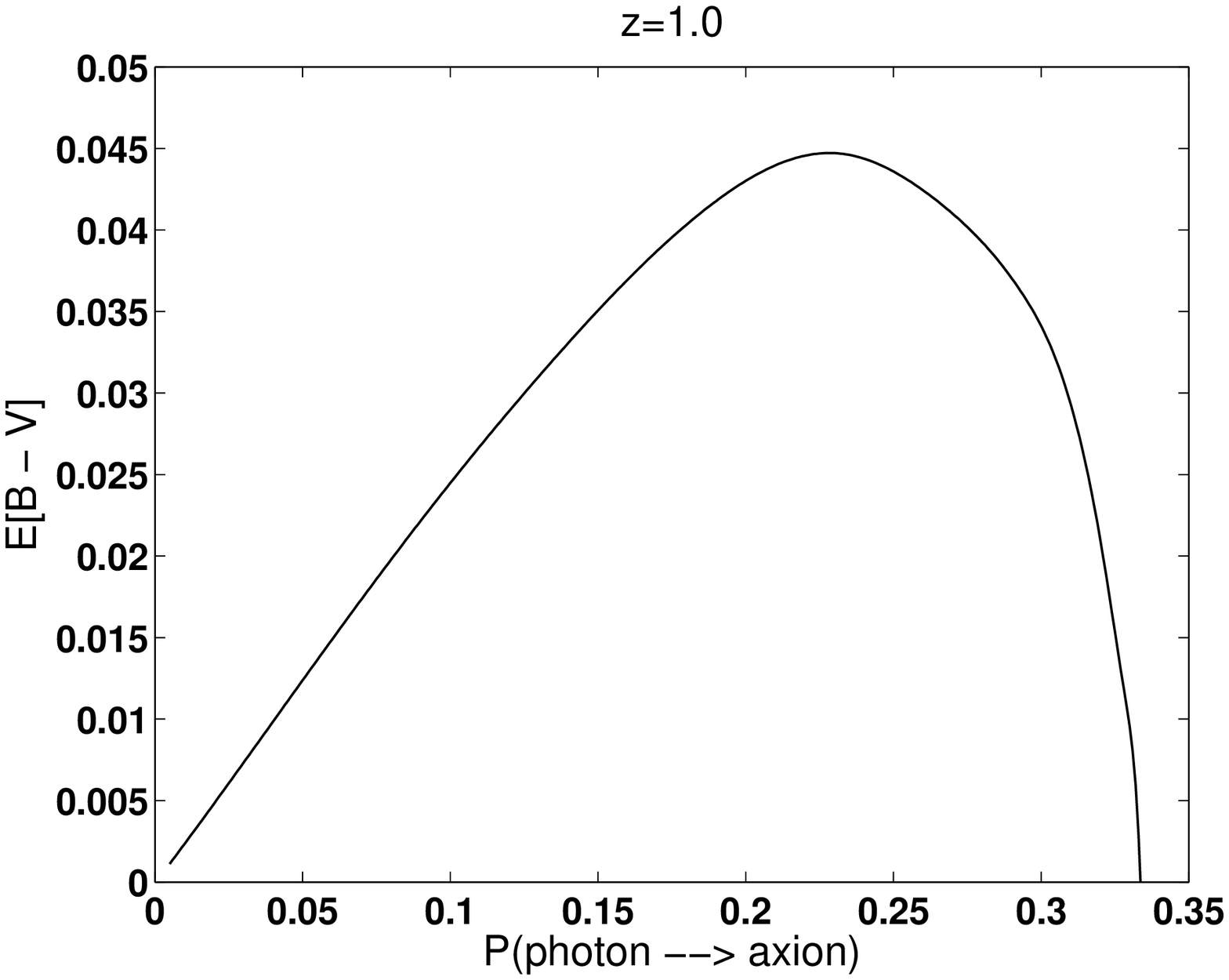}\cr
\end{tabular}    
\caption{Colour excess vs. dimming of photon flux due to mixing into axions for different values of $\eta$ at redshifts $z=0.5$ and $z=1$}
\label{achrom}
\end{figure}

\begin{figure}
\begin{tabular}{ccc}
\includegraphics[height=6cm,width=7.5cm]{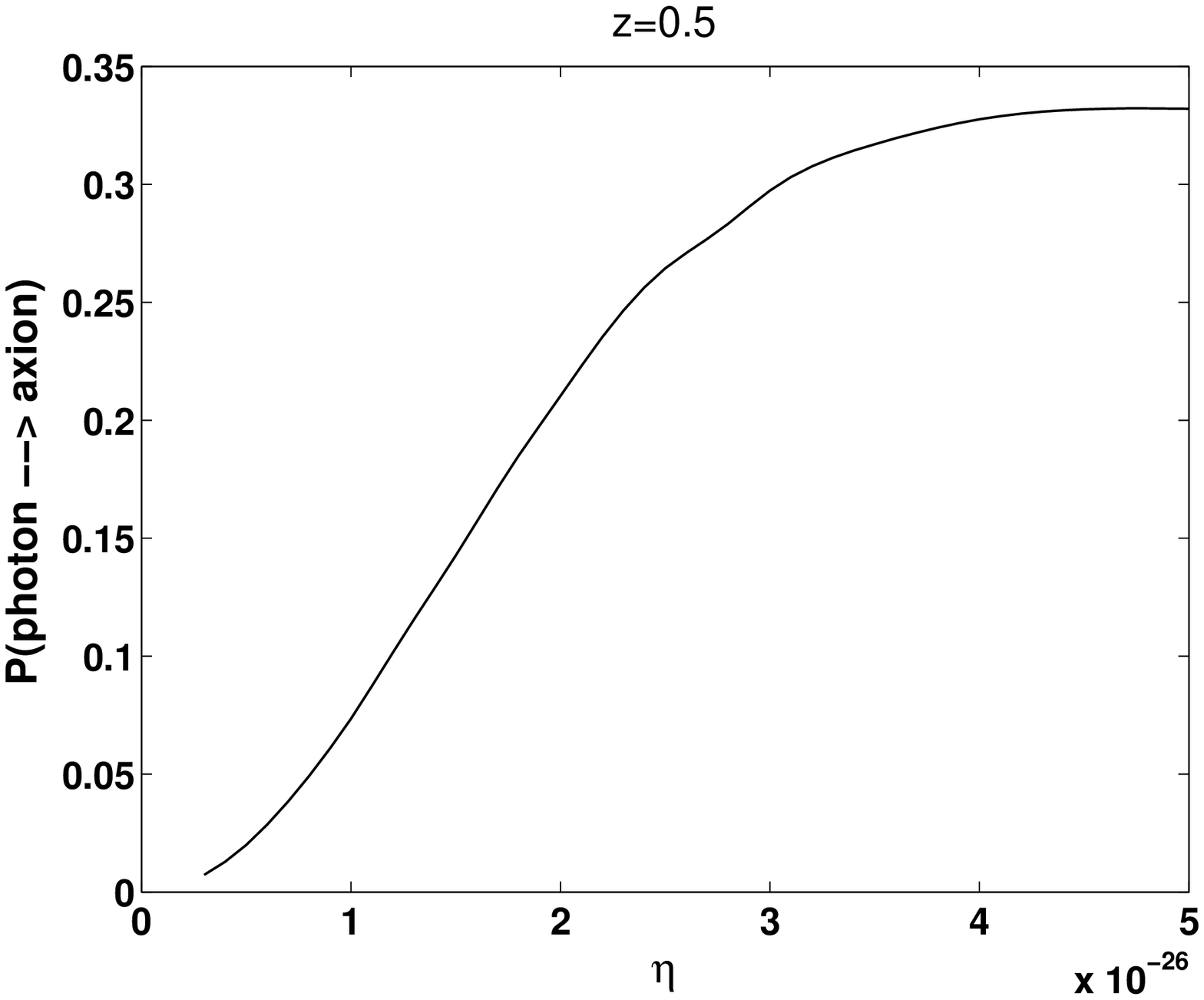} 
&\hspace{1.0cm}&
\includegraphics[height=6cm,width=7.5cm]{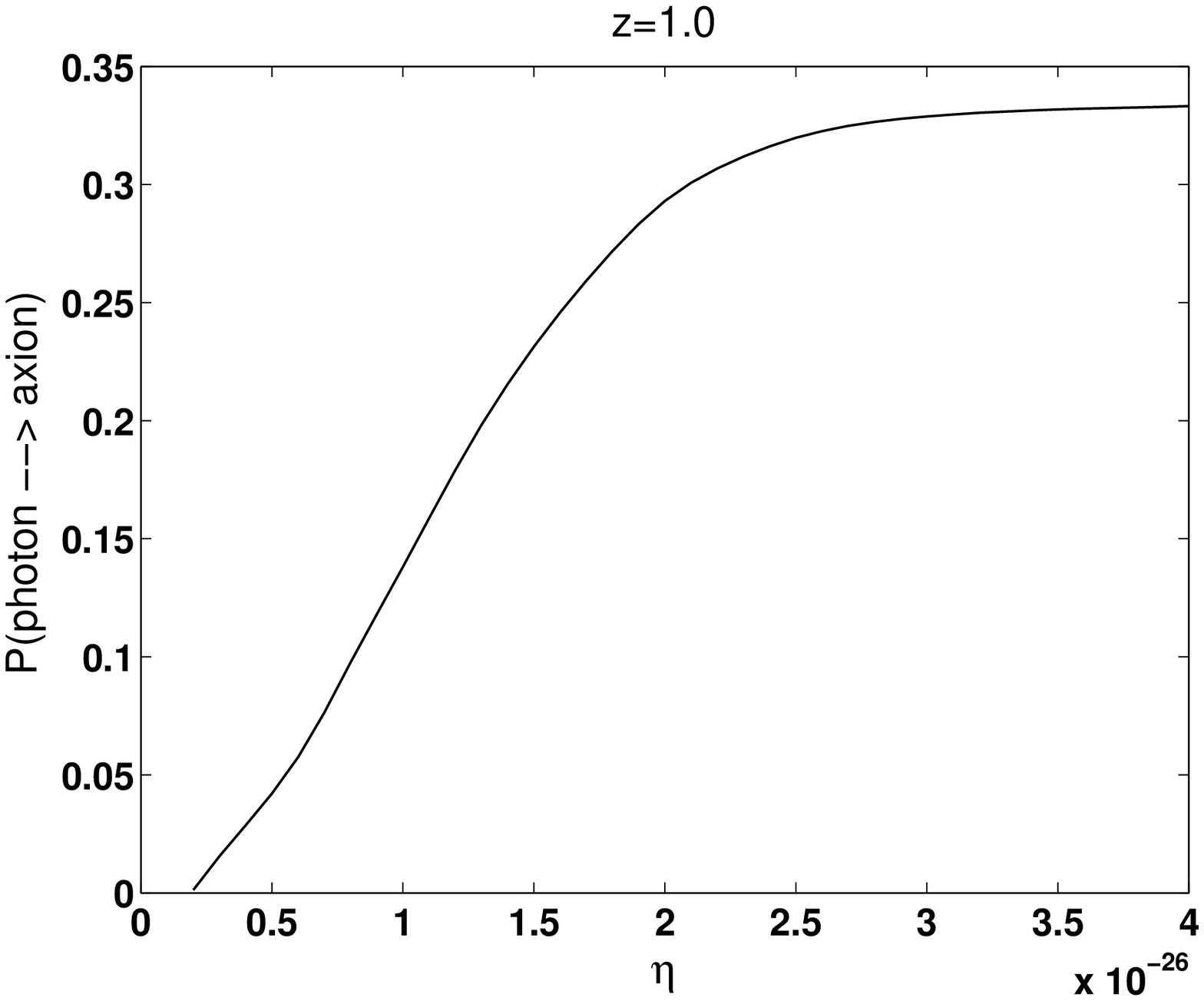}\cr
\end{tabular}    
\caption{Dimming of photon flux vs. $\eta$ at redshifts $z=0.5$ and $z=1$}
\label{dim}
\end{figure}

The colour excess $E(B-V)$ is a parameter normally used to quantify the difference in the ratio of emitted and observed flux (denoted $F_{\rm emi}$ and $F_{\rm obs}$ respectively) between the $B$ and $V$ bands due to intervening matter.  It therefore gives an indication of the frequency dependence of any dimming between the observer and the source.  The colour excess is defined as
\begin{equation}
E(B-V)=-2.5 \ {\rm log} \left(\frac{F_{\rm obs}(B)}{F_{\rm emi}(B)}\frac{F_{\rm emi}(V)}{F_{\rm obs}(V)}\right).
\end{equation}  
Supernova observations have placed an upper limit upon this quantity of $E(B-V)\leq 0.03$ \cite{sndim} so we must try to vary $\eta$ to obtain enough dimming in order to explain the supernova results whilst staying below this colour bound.  Figure \ref{achrom} plots the quantity $E(B-V)$ against the dimming parameter $\eta$ based upon our simulations using the background cosmology and density probability function described earlier.  This figure shows that the maximum amount of permitted dimming at a redshift of $z=1$ which would still satisfy the colour bound is about 12 \%.  In figure \ref{dim} we have plotted photon dimming vs. the parameter $\eta$.  A dimming of 12 \% at $z=1$ indicates a value for this parameter of $\eta\sim 1\times 10^{-26}\rm cm^{-1}$ which also suggests a dimming of only about 8\% at $z=0.5$.  A larger dimming which would better account for the data at $z\sim 0.5$ is $\eta=1.5\times 10^{-26}\rm cm^{-1}$ although one might then have already expected to see supernovae which violate the colour bound.

\subsection{Polarisation}

Similarly we can investigate the evolution of light from a polarised source (see also \cite{sikivie,jain} for related previous work).  The magnetic fields in between us and the polarised source will not of course be aligned with the polarisation axis of the source.  Photons will therefore be able to mix into axions and axions into photons and the net angular momentum of the flux will not be conserved since the mixing between the photons and the axion field transfers angular momentum to and from the surrounding magnetic field.
By carefully analysing the mixing we are therefore able to find out the way a highly polarised source loses its polarisation as the photons propagate through the intergalactic medium.  As in the case of dimming, it is important for the magnetic field orientation and the direction of propagation of the photons to be uncorrelated in order for information concerning the polarisation to be lost in this way.  However, since we expect there to be a large number of cells with randomly oriented B-fields in between us and the source, this does in fact occur.

Let us consider a completely (100\%) linearly polarised source at a particular redshift.  As before, we call the two axes perpendicular to the direction of photon propagation and each other $x$ and $y$.  We choose to align $x$ with the polarisation axis of the source which corresponds to an initial density matrix of the form
\ben
\rho_{initial}=\left(
\begin{array}{ccc}
1&0&0\\
0&0&0\\
0&0&0\end{array}
\right)
\een
and then calculate the mixing of the photons from such a source.  As these photons pass through the inhomogeneous matter distribution in our simulations we can define a polarisation ratio $\Pi$ at a time $t_n$ after $n$ cells as
\begin{equation}
\Pi(t_n)=\left|\frac{\rho_{xx}(t_n)-\rho_{yy}(t_n)}{\rho_{xx}(t_n)+\rho_{yy}(t_n)}\right|
\end{equation}
which tells us how much the linear polarisation of the light will have been diminished in reaching us.

\begin{figure}
\begin{center}
\includegraphics[height=8cm,width=10cm]{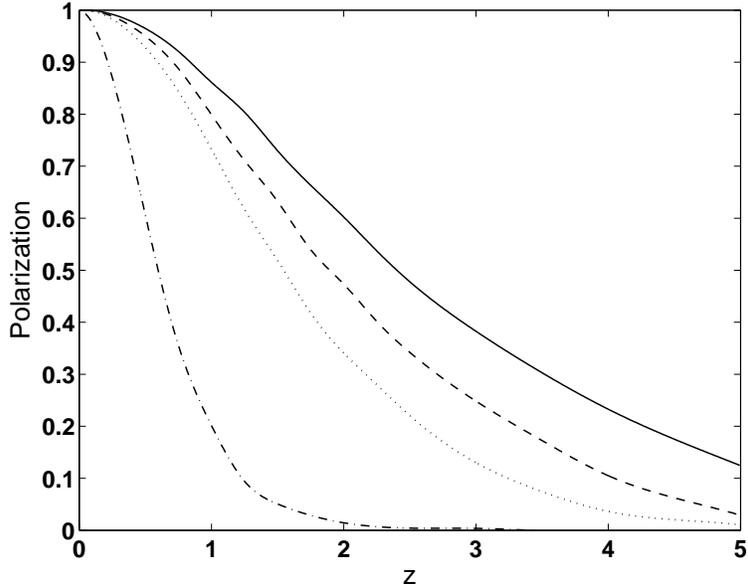}
\caption{Observed polarisation ratio $\Pi$ of completely polarised sources located at different redshifts $z$ and seen at different $\lambda$.  Solid line is B band (440 nm), dashed line is U band (360 nm), dotted line is edge of UV spectrum (300 nm) and dot-dash line is 100 nm. Dimming parameter $\eta=1.5\times 10^{-26}{\rm cm}^{-1}$.}
\label{pol}
\end{center}
\end{figure}

Figure $\ref{pol}$ shows the observed polarisation as seen on earth of a completely polarised source located at a redshift $z$ assuming the light from the source has been diluted by photon-axion mixing along the way with a dimming parameter of $\eta=1.5\times 10^{-26}{\rm cm}^{-1}$.  We have considered mixing of light from four different parts of the spectrum, the B and U bands in the optical, the boundary between the 'visible' and the ultraviolet window (300nm) and well into the ultraviolet (100nm).  Any astronomically observed source with redshift and polarisation values which lie above these curves will rule out this degree of mixing.  

Table \ref{poltab} contains a few results of polarimetric observations of high redshift active galactic nuclei\footnote{Table \ref{poltab} does not represent an exhaustive literature search so it is entirely possible that data exists which could constrain the parameters more strictly.}.  The values listed in this table are compatible with this value of $\eta$, so it would not be necessary to reduce the amount of dimming in order to accommodate these observations.  Consequently, to our knowledge, polarimetric observations do not yet rule out cases where the dimness of supernovae can be explained by axions.
\begin{table*}
\begin{center}
\caption[]{\label{poltab} Some observations of highly polarised high redshift active galactic nuclei. $\lambda_{obs}$ is the central wavelength of the light used to obtain $\Pi$ in the rest frame of the telescope.}\vspace{0.5cm}
\begin{tabular}{|c|c|c|c|c|}\hline
Object          &Redshift(z)    &$\lambda_{obs}$(nm)	&$\Pi (\%)$    &Reference \\ \hline
4C+03.24        &3.56           &604	&11.3$\pm$3.9  	&\cite{pol2} \\ \hline 
4C+00.54	&2.37		&443	&11.9$\pm$2.6	&\cite{499L21}\\ \hline
0211-122	&2.34		&435	&19.3$\pm$1.1	&\cite{pol2} \\ \hline
FSC 10214+4724  &2.28           &416	&26  $\pm$2 	&\cite{pol1}\\ \hline
0823-223	&$>$0.91		&350	&16.8$\pm$0.84	&\cite{nordic}\\ \hline
1522+101	&1.32		&166	&4.7$\pm$1.9&\cite{hstuvpol}\\ \hline
\end{tabular}
\end{center}
\end{table*}

However, the polarisation effect is extremely frequency dependent, and figure $\ref{pol}$ shows that we do not expect to see any highly polarised high redshift sources in the ultra-violet part of the spectrum ($\lambda\sim$100nm) with values of $\eta$ which could lead to significant dimming.  UV Polarimetry of high redshift objects must be done from space and table $\ref{poltab}$ includes observations using the Hubble space telescope faint object spectrophotometer \cite{hstuvpol}.  Such observations should also be possible with the new instruments on the Hubble space telescope.  

\section{Discussion}

In this paper we have considered the dimming of light from high redshift supernovae due to mixing between photons and a pseudoscalar axion field in the intervening intergalactic plasma.  We have tried to simulate this process in a more detailed way than previous studies by taking a realistic density probability distribution for the electron density.  We have also included the effect of a fluctuating magnetic field due to it being frozen into the background plasma.  

The presence of such an axion does not completely solve the problem of the missing $\sim 70\%$ of the energy of the universe.  Rather we aimed to show that such a mixing might make it possible to explain the missing dark energy of the universe via an alternative source of stress energy other than a cosmological constant.  As a candidate for this energy we considered the case of a string network with equation of state $\omega_S=-1/3$ and calculated the combination of axion-photon coupling and magnetic field that would lead to enough dimming.  

Assuming a primordial origin for the intergalactic magnetic fields we found no values of these parameters which could explain the dimming without violating the bound on the frequency dependence of this dimming already calculated by the observational supernovae groups.  However, the colour bound that we observed was only a factor of about 50\% larger than that observed, not many orders of magnitude, and there is a small inherent error expected in our analysis due to deviations from our assumed behaviour of the variance parameter $\sigma(z)$ in the density probability function (\ref{Pdelta}) which will become greater at high redshifts.  We found that for $\eta=1.5\times 10^{-26}{\rm cm}^{-1}$ we obtained nearly enough dimming and only marginally violated he colour bound.  We therefore have adopted this as our optimal value of $\eta$.

One of the main conclusions of this work is therefore that a photon-axion coupling of the sort discussed in \cite{Csaki:2001yk} leads to very particular predictions as to the redshift dependence of dimming at different frequencies.

If one considers an alternative origin for the intergalactic magnetic fields other than their primordial production it will be possible to change the bounds obtained in this paper since the correlation between magnetic field strength and electron density used in our calculations will be modified.

We also calculated the expected reduction in the polarisation of light from high redshift objects due to photon-axion mixing.  After a brief literature search we were unable to find any observations of high redshift AGN with large enough polarisations to be at odds with a dimming of $\eta=1.5\times 10^{-26}{\rm cm}^{-1}$.  However, we find the frequency dependence of this effect to be so large that we believe such a dimming would lead to no observed sources at redshift $z\sim 1.5$ with UV (100 nm) polarisation as high as $10\%$.  Any distribution of magnetic fields which could lead to significant dimming of the optical flux coming from a distant object should also lead to a large reduction in any polarisation of the UV photons coming from that object.  This prediction is independent of the detailed history of cosmic magnetic fields.  It should be possible for this issue to be further investigated by making space based polarimetric observations.

\section*{Acknowledgements} We are grateful for comments on the first version of this paper by Csaba Csaki, Nemanja Kaloper and John Terning.  We are also grateful to the referee for persuading us to implement the full density matrix analysis.  MF is funded by an IISN grant and the IUAP program of the Belgian Federal Government.

\end{document}